\documentclass{PoS}
\usepackage{psfrag}

\title{Lattice Quantum Gravity - an Update}

\ShortTitle{Lattice Quantum Gravity}

\author{Jan Ambj\o rn\\
        The Niels Bohr Institute, Copenhagen University, 
        Blegdamsvej 17, DK-2100 Copenhagen \O , Denmark\\
        E-mail: \email{ambjorn@nbi.dk}}

\author{Jerzy Jurkiewicz\\
        Institute of Physics, Jagellonian University, Reymonta 4, 
        PL 30-059 Krakow, Poland\\
        E-mail: \email{jurkiewicz@th.if.uj.edu.pl}}

\author{\speaker{Renate Loll}
\\
        Institute for Theoretical Physics, Utrecht University, Leuvenlaan 4, NL-3584 CE Utrecht, The Netherlands.\\
        E-mail: \email{r.loll@uu.nl}}

\abstract{We advocate lattice methods as the tool of choice to constructively
define a background-independent theory of Lorentzian quantum gravity
and explore its physical properties in the Planckian regime.
The formulation that arguably has most furthered our understanding of
quantum gravity (and of various pitfalls present in the nonperturbative sector)
uses dynamical triangulations to regularize the nonperturbative path integral 
over geometries. Its Lorentzian version in terms of Causal Dynamical 
Triangulations (CDT) -- in addition to having a definite quantum
signature on short scales -- has been shown to reproduce important 
features of the classical theory on large scales. This article recaps the most
important developments in CDT of the last few years for the physically
relevant case of four spacetime dimensions, and describes its status quo 
at present.}

\FullConference{The XXVIII International Symposium on Lattice Field Theory, Lattice2010\\
		June 14-19, 2010\\
		Villasimius, Italy}

\begin{document}

\section{Quantum gravity and the lattice}\label{sec1}

Gravity remains the only fundamental interaction which we have not yet been 
able to formulate and understand as a full quantum theory. Even basic issues remain
open, for example, whether gravity at the most fundamental level will be truly 
unified with the weak, strong and electromagnetic interactions, or have more of 
a separate status, in the spirit of the classical theory. Research in quantum gravity
is driven by a number of simple, but profound questions: What are the quantum
origins of space, time and our universe? What is the microstructure of spacetime, 
and can it {\it explain} macroscopic gravitational interactions and perhaps even the 
universe's observed large-scale structure?
Are ``space'', ``time'' and ``causality'' fundamental or emergent concepts in a 
setting where spacetime geometry is allowed to undergo large quantum fluctuations?

Apart from its appeal as a white spot on the map of our understanding of 
fundamental high-energy physics,
the specific reason why this topic is of inherent interest to the lattice community is 
the apparent need in quantum gravity
for nonperturbative methods to model and understand the relevant Planck-scale 
physics. Lattice and Monte Carlo techniques, adapted to systems of dynamical 
geometry (such as gravity), provide powerful tools for addressing such issues. 
For low-dimensional systems of quantum geometry the validity and usefulness
of such methods has been demonstrated long ago, and reviewed at 
previous lattice conferences under headings like 
``lattice gravity and random surfaces'' \cite{latticereviews}. 
Similar techniques can be applied to fully-fledged four-dimensional
quantum gravity, but the situation here is less clear-cut, which is not surprising 
in view of our limited understanding of this theory. As will be
described in what follows, attempts are
under way to {\it define} quantum gravity as the scaling limit of
a specific statistical system of dynamical geometry. For the physically relevant
case of four spacetime dimensions, the only way we can  
currently study the existence and properties of this nonperturbative limit is
via lattice methods. In other words, despite the fact that 
quantum gravity is at a much earlier stage of theory building,
compared with a theory like QCD, numerical methods -- in conjunction with 
analytical and theoretical modelling -- can be used in a profitable way
to explore what this theory may be. 
This also implies that the more foundational aspects of theory development are 
currently at least
on a par with purely simulation-technical aspects, like improving efficiency or
increasing the lattice size. 

At this stage, the only points of reference and comparison for lattice
quantum gravity are alternative and 
(likewise incomplete) nonperturbative formulations in the continuum. In addition, because 
of the requirement of covariance, there are also considerable 
challenges in defining and evaluating observable quantities, which can be used to 
characterize the physical properties of the theory.
The specific candidate theory of quantum gravity described below arises
from a confluence of ideas from general relativity (in particular, gravity-specific
properties like dynamical geometry and background independence),
high-energy physics (in particular, the use of path integral and
renormalization group methods), and, equally crucially, lattice field theory.  
This approach of ``{\it Quantum Gravity from Causal Dynamical Triangulations (CDT)}''
was last reported on during plenary talks at Lattice 2000 and 2001 \cite{latticecdt},
when the formulation was still in its infancy, and far from deriving results in the
physically interesting case of four dimensions. The remainder of this presentation
constitutes a brief progress report on the many interesting developments that
have taken place since then, focussing on four-dimensional results. More extensive
recent reviews of the field can be found in \cite{cdtreviews}.

\section{Causal Dynamical Triangulations 101}\label{sec2}

Viewed from a larger perspective, CDT quantum gravity is an outright conservative
approach in the sense of relying exclusively on standard quantum field-theoretic tools
and principles, applied to the situation where spacetime is not regarded as fixed, but
itself part of the nonperturbative dynamics. It builds on techniques 
which have been well tested in the study of systems of random surfaces and 
Euclidean\footnote{Instead of space{\it times} with Lorentzian signature, Euclidean gravity 
works with purely spatial geometries, which do not have a notion of time or causality.
Euclidean gravity was a popular starting point for cosmological path integrals back in
the 1970s and '80s and (usually for reasons of simplification) 
continues to be used in some path integral formulations of full gravity.} 
models of quantum gravity, and does not invoke or presently require ``exotic'' 
ingredients like strings, loops, branes, extra dimensions or new symmetries. 
By contrast, it is an approach with few free parameters, whose outcomes are
by construction robust. This means that if it can be shown to lead to a viable
theory of quantum gravity, the theory will be reasonably unique. On the other
hand, if in the future it produces results which are inconsistent (for example, 
because its
classical limit is in contradiction with Einstein's general relativity), it will be
difficult to fix this by twiddling with the parameters of the model. 
 
Since dynamically triangulated models of
quantum gravity are amenable to numerical methods, which can and do produce
numbers and results, the above considerations are not merely of a theoretical 
nature, as is
illustrated by the fate of the Euclidean precursor of CDT. This candidate
theory of four-dimensional quantum gravity generated considerable excitement
in the early 1990s before it was understood gradually that it suffers from fatal 
degeneracies, which prevent the emergence of macroscopic, classical spacetimes 
of dimension four. Numerical simulations were crucial in bringing about
this result (for a summary of these developments, as well as a complete
bibliography, see \cite{livrev}). This illustrates 
that the presence of explicit computational consistency checks, combined with a small
number of free parameters and a high degree of universality (independence of
the continuum theory of the details of the lattice discretization) means that in
practice quantum gravity theories from dynamical triangulations can be falsified. 
Despite being a hallmark of any good 
physical theory, falsifiability has become somewhat of a rarity
in more speculative areas of high-energy theory, including quantum gravity.
What we would like to emphasize here is the importance -- in the absence of any
direct probes of Planck-scale physics -- of ``computational experiments'' in
providing criteria for the viability of candidate theories for quantum gravity.

In technical terms, quantum gravity from causal dynamical triangulations is
a nonperturbative implementation of the gravitational path integral. It has 
already passed several nontrivial tests and has produced unprecedented results,
as will be described below.
In the process, it also has highlighted a number of unexpected features (and
pitfalls) due to the nonperturbative nature of the construction, which permits
large quantum fluctuations on small scales. 

The
idea of constructing a nonperturbative gravitational path integral which captures   
Lorentzian, causal properties of the spacetimes to be summed over goes back
to a paper from 1998 \cite{al}, where it was also demonstrated by explicit, analytic
computation that the idea works in two dimensions and produces a result
distinct from previous Euclidean models of 2d quantum gravity. The first
results for the physical, four-dimensional theory were published in 2004
\cite{ajl-prl}.  

As a warm-up, consider the path integral for a nonrelativistic particle of mass 
$m$ in
one dimension, moving in a harmonic oscillator potential, and subject to
fixed boundary conditions $x_i=x(t_i)$ and $x_f=x(t_f)$ at some initial and
final times $t_i$ and $t_f$. The corresponding path integral describes the
transition amplitude from $x_i$ to $x_f$ as a superposition of amplitudes
${\rm exp}\ iS[x(t)]$ of all possible particle trajectories with the given
boundary conditions, where $S[x(t)]$ is the classical action of the entire
path $x(t)$.
\begin{figure}[t]
\psfrag{a}{{\bf{\LARGE $\tau_i$}}}
\psfrag{b}{{\bf{\LARGE $\tau_f$}}}
\psfrag{c}{{\hspace{-20pt}\bf{\LARGE $x_i$}}}
\psfrag{d}{{\hspace{-20pt}\bf{\LARGE $x_f$}}}
\psfrag{t}{{\bf{\LARGE $\tau$}}}
\psfrag{x}{{\bf{\LARGE $x$}}}
\psfrag{u}{{\bf{\LARGE $a$}}}
\centerline{\scalebox{0.8}{\rotatebox{0}{\includegraphics{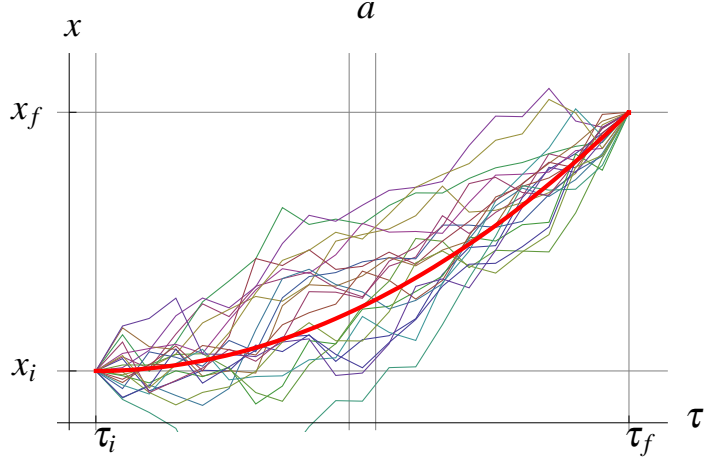}}}}
\caption{Sample paths or ``histories'' $x(\tau)$ from a regularized version of
the path integral of the 
nonrelativistic particle in imaginary time $\tau:=-i t$, 
generated by a Monte Carlo simulation. The total
time interval $T=\tau_f-\tau_i$ has been subdivided into time steps of length $a$,
and the trajectories are piecewise linear.
The average path $\langle x(\tau )\rangle$ is indicated by the central fat line.}
\label{line}
\end{figure}
The superposition gives rise to an ``average path'', the expectation value
$\langle x(t)\rangle$ in the given ensemble (c.f. Fig.\ \ref{line}, which shows
a sample of paths in Euclidean,
imaginary time $\tau:=-i t$, generated by a Monte Carlo simulation). 
The typical size of the deviation $\delta x(\tau)$ 
of a general history $x(\tau)=\langle x(\tau)\rangle +\delta x(\tau)$ from this expectation 
value can be computed explicitly, yielding
\begin{equation}
\langle \delta x(\tau)^2\rangle = \frac{\hbar}{2 m\omega}\frac{\cosh \omega T-
\cosh \omega (T-2 \tau)}{\sinh \omega T},
\end{equation}
where $\omega$ is the oscillator's frequency, 
$T=\tau_f-\tau_i$ is the total length of the time interval considered, and
the time $\tau$ runs from $\tau_i=0$ to $T$. 
Anticipating a similar
quantum superposition in gravity, where each ``path'' will 
represent a curved spacetime,
we will be interested both in the ``average universe'' and deviations from it. In
this case, the scale of quantum fluctuations of some linear distance $x$ is
expected to be $\langle |\delta x |\rangle\propto \sqrt{\hbar G}$ instead of
the $\langle |\delta x |\rangle\propto \sqrt{\hbar /m\omega}$ of the particle case,
with $G$ denoting Newton's constant. 

Quantum gravity from causal dynamical triangulations is a
nonperturbative and background-independent realization of the formal 
{\it gravitational 
path integral} (a.k.a. the ``sum over histories'') on a differential manifold $M$,
\begin{equation}
Z(G,\Lambda)=
\int\limits_{ {\cal G}({\rm M})=\frac{\rm Lor(M)}{\rm Diff(M)}}
{\cal D}[g_{\mu\nu}]\ {\rm e}^{iS^{\rm EH}[g_{\mu\nu}]}, \;\;\;
\label{gravint}
\end{equation}
where $S^{\rm EH}$ is the four-dimensional Einstein-Hilbert action, 
$\Lambda$ the cosmological constant, and the path integral is to be taken 
over all spacetimes $[g_{\mu\nu}]\in {\cal G}(M)$ (Lorentzian 
metrics $g_{\mu\nu}$ modulo diffeomorphisms), with specified 
boundary conditions.  
In other words, each path is now a four-dimensional, curved spacetime 
geometry $[g_{\mu\nu}]$,
which can be thought of as a three-dimensional, spatial geometry
developing in time. The weight associated with each $[g]\in {\cal G}(M)$ is given
by the Einstein-Hilbert action
\begin{equation}
S^{\rm EH}[g]=\frac{1}{16\pi G}\int d^4x \sqrt{-\det g}(R[g,\partial g,
\partial^2g]-2\Lambda).
\label{contlang}
\end{equation} 
To evaluate this quantum field-theoretic path integral,
one proceeds in close analogy with the path-integral quantization of the 
nonrelativistic particle described above. The latter is
defined as the continuum limit of a regularized sum over paths,
where the contributing ``virtual'' paths are taken from an ensemble of piecewise
straight paths, with the time interval $a$ for each step going to zero in the limit.
The method of CDT turns the corresponding gravitational path integral 
(\ref{gravint}) into a well-defined regularized and
finite expression, which can be evaluated and whose continuum limit 
can be studied systematically \cite{ajl1}.
The CDT prescription consists in representing the 
space ${\cal G}(M)$ of all Lorentzian spacetimes 
in terms of a set of triangulated, 
piecewise flat (ie. piecewise Minkowskian) manifolds\footnote{Unlike in the 
particle case,
there is no embedding space; all geometric spacetime data are defined 
intrinsically, just like in the classical theory.}. 

The idea of approximating curved spacetimes by much simpler, 
triangulated objects was introduced in the classical
theory of General Relativity by Regge \cite{regge}, and first applied in the
quantum context in a seminal paper by Ro\v cek and Williams \cite{rw}. 
Note that the objectives of the classical and quantum theories differ
significantly: in the former, one usually wants to approximate a given, classical
solution to the Einstein equation locally as well as possible. By contrast, when using
such geometries in the path integral, one wants to 
approximate the space of {\it all} geometries. It should be pointed out that
just like in the 
particle case, where the path integral in the continuum limit is dominated by
nowhere differentiable paths, typical geometries contributing 
to the gravitational path integral also turn out to be highly nonclassical.  

The geometry of the triangulated manifolds is almost everwhere flat and therefore
trivial, and can carry curvature in a delta function-like manner
only at its two-dimensional subsimplices (the triangles), where three or more
four-simplices meet. This regularization in terms of dynamical lattices
implies a vast truncation of the number of
degrees of freedom, from the local field tensor $g_{\mu\nu}(x)$ to a discrete set
of edge lengths for the four-simplices, plus the information of which pairs
of simplices are glued together pairwise.

\begin{figure}[t]
\centerline{\scalebox{0.5}{\rotatebox{0}{\includegraphics{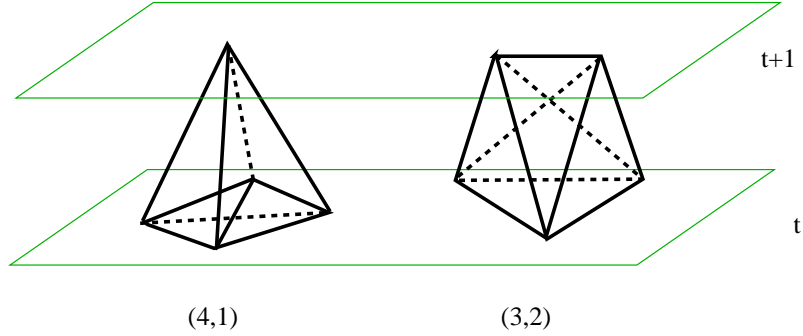}}}}
\caption{The two fundamental building blocks of CDT are four-simplices with
flat, Minkowskian interior. They are spanned by spacelike edges, 
which lie entirely within spatial slices of constant proper
time $t$, and timelike edges, which interpolate between adjacent slices of
integer time. A building block of type $(m,n)$ has $m$ of its vertices in
slice $t$, and $n$ in slice $t+1$.
}
\label{buildingblocks}
\end{figure}

For the purposes of causal dynamical triangulations,
the simplicial approximation ${\cal G}_{a,N}$ of $\cal G$ contains
all simplicial manifolds $T$ obtained from gluing together at most 
$N$ four-dimensional,
triangular building blocks of typical edge length $a$, with $a$ again playing
the role of a UV cut-off (see Fig.\ \ref{buildingblocks}). 
What makes the construction {\it causal} is the fact that the gluing of the
Minkowskian four-simplices respects a global notion of (proper) time, 
akin to the requirement of global hyperbolicity usually imposed in classical
gravity. The regularized gravitational path integral in CDT is then given by  
\begin{equation}
\label{discretesum}
Z^{\rm CDT}_{a,N}= \sum_{{\rm triangulated\ causal} \atop 
{\rm spacetimes} \, T\in{\cal G}_{a,N}}\frac{1}{C_T}{\rm e}^{i S^{\rm Regge}[T]},
\end{equation}
where $S^{\rm Regge}$ is the Regge version of the Einstein-Hilbert action 
associated with the simplicial spacetime $T$, and  
$C_T$ denotes the order of its automorphism group (see \cite{ajl-rec,desitter1}
or the recent reviews \cite{details4d} for an 
explicit expression of $S^{\rm Regge}$ as well as other construction details).
The discrete volume $N$ acts as an infrared cutoff. 
We still need to consider a suitable continuum or scaling limit 
\begin{equation}
\label{discretelim}
Z^{\rm CDT}:=\lim_{N\rightarrow\infty\atop a\rightarrow 0} Z^{\rm CDT}_{a,N}
\end{equation}
of (\ref{discretesum}), while renormalizing the original bare coupling constants
of the model, in order to arrive at a theory of quantum gravity.\footnote{Note
that the existence of
a physically meaningful limit is not automatic, but something that needs to be
shown.} The two limits in (\ref{discretelim}) are
usually tied together by nominally keeping fixed a physical four-volume
$V_4:=a^4 N$. 
In order to make the evaluation of $Z$ amenable to Monte Carlo simulations,
one still needs to convert the sum over complex amplitudes to a sum over
real Boltzmann weights. Despite the fact that no suitable Wick rotation is
known for arbitrary curved metrics, such a prescription fortunately does exist
for the causal triangulations under consideration \cite{ajl1}.
As is familiar from lattice field theory, one then takes $a\rightarrow 0$, 
such that the individual discrete building blocks shrink to zero. This should
be contrasted with some other approaches to quantum gravity, which
postulate the existence of fundamental discreteness at the Planck scale, and
consequently identify the lattice spacing $a$ with the Planck length 
$\ell_{\rm Pl}$. 
In this case one never takes a continuum limit $a\rightarrow 0$, which has
the disadvantage that the quantum dynamics at the Planck scale is {\it not}
universal and has a large degree of arbitrariness.
In CDT applications, since the limit $a\rightarrow 0$ can 
in practice never be reached
on a finite lattice, one must make sure that $a$ is always much smaller than 
the scale at which one is trying to extract physical results.

Let us summarize the key features of the construction scheme thus introduced.
Unlike what is possible in the continuum theory, the path integral 
(\ref{discretesum}) is defined directly on the physical configuration space of
{\it geometries}. It is nonperturbative in the sense of including 
geometries which are ``far away'' from any classical solutions, and it is
background-independent in the sense of performing the sum ``democratically'',
without distinguishing any given geometry (say, as a preferred background). 
However, these attractive properties of the regularized path integral are only 
useful because \textit{we are able to evaluate $Z^{\rm CDT}$ quantitatively},
with an essential role being played by Monte Carlo simulations. These, together
with the associated finite-size scaling techniques \cite{newmanbarkema}, have
enabled us to extract information about the nonperturbative, 
strongly coupled quantum
dynamics of the system, which is currently not accessible by analytical
methods, neither in this nor any other approach to quantum gravity. 
This mirrors the role played by lattice simulations in determining
the nonperturbative behaviour of QCD (although one should keep in mind
that the latter is a theory we 
already know {\it much} more about than quantum gravity). 

As far as we are aware, CDT is the only 
nonperturbative approach to quantum gravity 
which has been able to dynamically generate its own, physically realistic
background from nothing but quantum fluctuations. More than that, because
of the minimalist set-up and the methodology used (quantum field theory
and critical phenomena), the results obtained are robust in the sense of
being largely independent of the details of the chosen regularization procedure 
and containing few free parameters. As we already pointed out in Sec.\ \ref{sec2}
above, it is therefore one of the 
rare instances of a candidate theory of quantum gravity which can potentially
be falsified.
Our investigations of both the quantum properties and the classical limit
of this candidate theory are at this stage not sufficiently complete to provide 
conclusive evidence that we have found {\it the} correct theory of quantum gravity.
However, results until now have been unprecedented and very encouraging, and
have thrown up a number of nonperturbative surprises, some of which we will
summarize next.

\begin{figure}[t]
\centerline{\scalebox{0.45}{\rotatebox{0}{\includegraphics{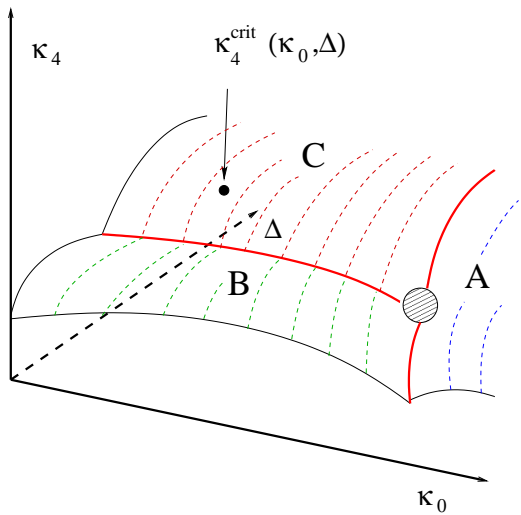}}}}
\caption{The phase diagram of Lorentzian quantum
gravity from CDT, with $\kappa_0$ and $\kappa_4$ denoting
the bare inverse Newton's constant and (up to an additive shift) the bare
cosmological constant. After fine-tuning to the subspace where 
the cosmological
constant is critical (tantamount to performing the infinite-volume limit), there are
three phases: A and B (the Lorentzian 
analogues of the degenerate branched-polymer and crumpled phases of
the Euclidean approach), and a {\it new}
phase C, where an extended, four-dimensional universe emerges. The parameter
$\Delta$ in CDT parametrizes a finite relative scaling between space- and
time-like distances which is naturally present in the Lorentzian case.}
\label{phases}
\end{figure}

\section{Key findings of CDT -- the phase diagram}\label{sec3}

One important lesson learned for nonperturbative
gravitational path integrals from CDT quantum gravity is that the ad-hoc
prescription of integrating over curved Euclidean {\it spaces} of metric signature 
(++++) instead of the physically correct curved Lorentzian space{\it times}
of metric signature ($-$+++) generally leads to inequivalent and (in $d\! =\! 4$)
incorrect results. Euclidean quantum gravity, as advocated
by Hawking and others \cite{eqg}, adopts the Euclidean version of
the path integral mainly for the technical reason of being able to 
use real weights $\exp (-S^{\rm eu})$ instead of the complex amplitudes
$\exp (i S^{\rm lor})$ in its evaluation. The same is done
in perturbative quantum field theory on flat Minkowski space, where
one can rely on the existence of a well-defined Wick rotation to 
relate correlation functions in either signature, an option that is {\it not} available
in continuum gravity beyond perturbation theory on a
Minkowski background.

CDT quantum gravity has given us 
the first explicit example of a nonperturbative
gravitational path integral (in a toy model of two-dimensional gravity \cite{al}) 
which is exactly
soluble and leads to distinct and inequivalent results, depending on whether
the sum over histories is taken over Euclidean spaces or Lorentzian spacetimes.
(More precisely, the latter are Euclidean spaces which are obtained by Wick rotation --
which {\it does} exist for the class of simplicial spacetimes under
consideration -- from Lorentzian spacetimes).  
Only those histories are summed over which 
possess a global time slicing with respect to which no spatial topology
changes are allowed to occur. After Wick rotation, this set constitutes a
strict subset of all Euclidean (triangulated) spaces. Note that general
Euclidean spaces possess no natural
notion of time or causality and in this sense branching 
in all directions is always present.

A crucial insight of CDT quantum gravity is that a similar result holds also in four
dimensions. The geometric degeneracy of the phases (in the sense of 
statistical systems) found in Euclidean dynamical triangulations and the
resulting absence of a good classical limit \cite{bialas,debakker}
can again in part be traced to the proliferation of branching 
``baby universes''. As demonstrated by the CDT 
results in \cite{ajl-prl,ajl-rec}, the requirement of microcausality (absence
of causality-violating points) of the individual path integral histories leads to
a different phase structure, compared with the previous Euclidean
approach. The breakthrough result of Lorentzian CDT is that its phase diagram
now possesses a third and {\it qualitatively new} phase, in which the universe on 
large scales is extended and four-dimensional (Fig.\ \ref{phases}), exactly
as required by classical General Relativity!
As indicated on the figure, to obtain an infinite-volume limit 
the bare cosmological constant $\kappa_4$ has
to be fine-tuned to the critical surface from {\it above}, since $\kappa_4 >
\kappa_4^{\rm crit}$ characterizes
the region where the (Euclidean) partition function $Z^{\rm CDT}$ exists and is finite. 

On the critical surface, 
phases A and B can be understood as Lorentzian analogues
of the two degenerate phases of the Euclidean models, and do not appear
interesting from a continuum point of view \cite{ajl-rec}. The new and
physically interesting phase -- more on which below -- is phase C. 
What is curious about the phase structure of four-dimensional CDT
quantum gravity is its resemblance with that of Ho\v rava-Lifshitz gravity
\cite{horava}, which has been spelled out further in \cite{lifshitzphases,horava-new}.
It gives rise to the intriguing conjecture that there may be a universal phase diagram
governing systems of higher-dimensional, dynamical geometry, and
accomodating a
variety of gravity theories, some of which may be anisotropic in space and time. 
Another question that arises is that of the order of the phase
transitions between the three phases, indicated by the red lines in Fig.\ \ref{phases}.
Their determination is numerically challenging, and a preliminary
investigation of the A-C transition in \cite{semilim} turned out inconclusive.
Some of these problems have now been overcome and new results on both 
the A-C and the B-C transition will appear in due course \cite{newphase}.

\section{Key findings of CDT -- the dynamical emergence of spacetime as
we know it}\label{sec4}
\begin{figure}[t]
\centerline{{\scalebox{0.4}{\rotatebox{90}{\includegraphics{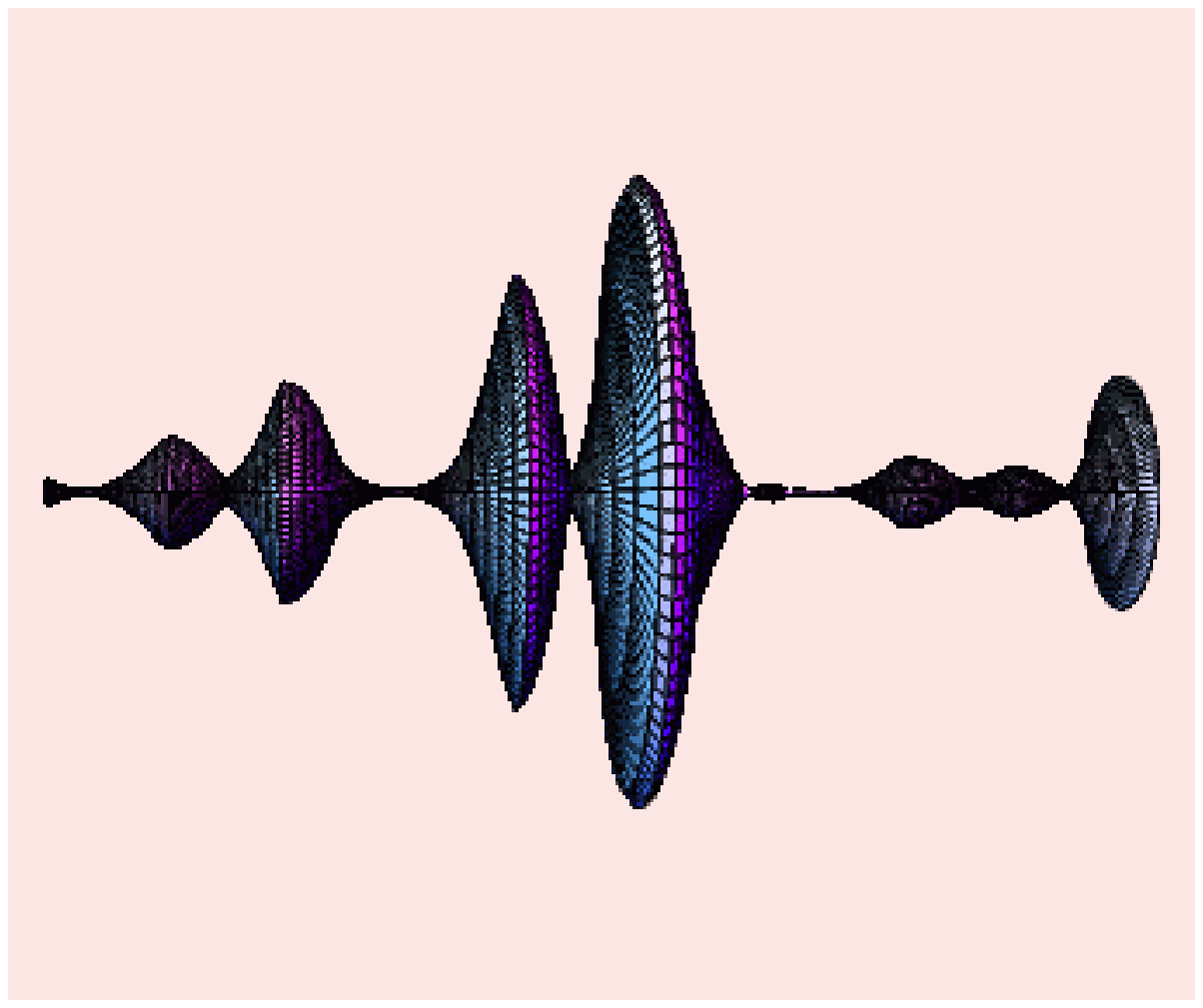}}}}
{\scalebox{0.4}{\rotatebox{90}{\includegraphics{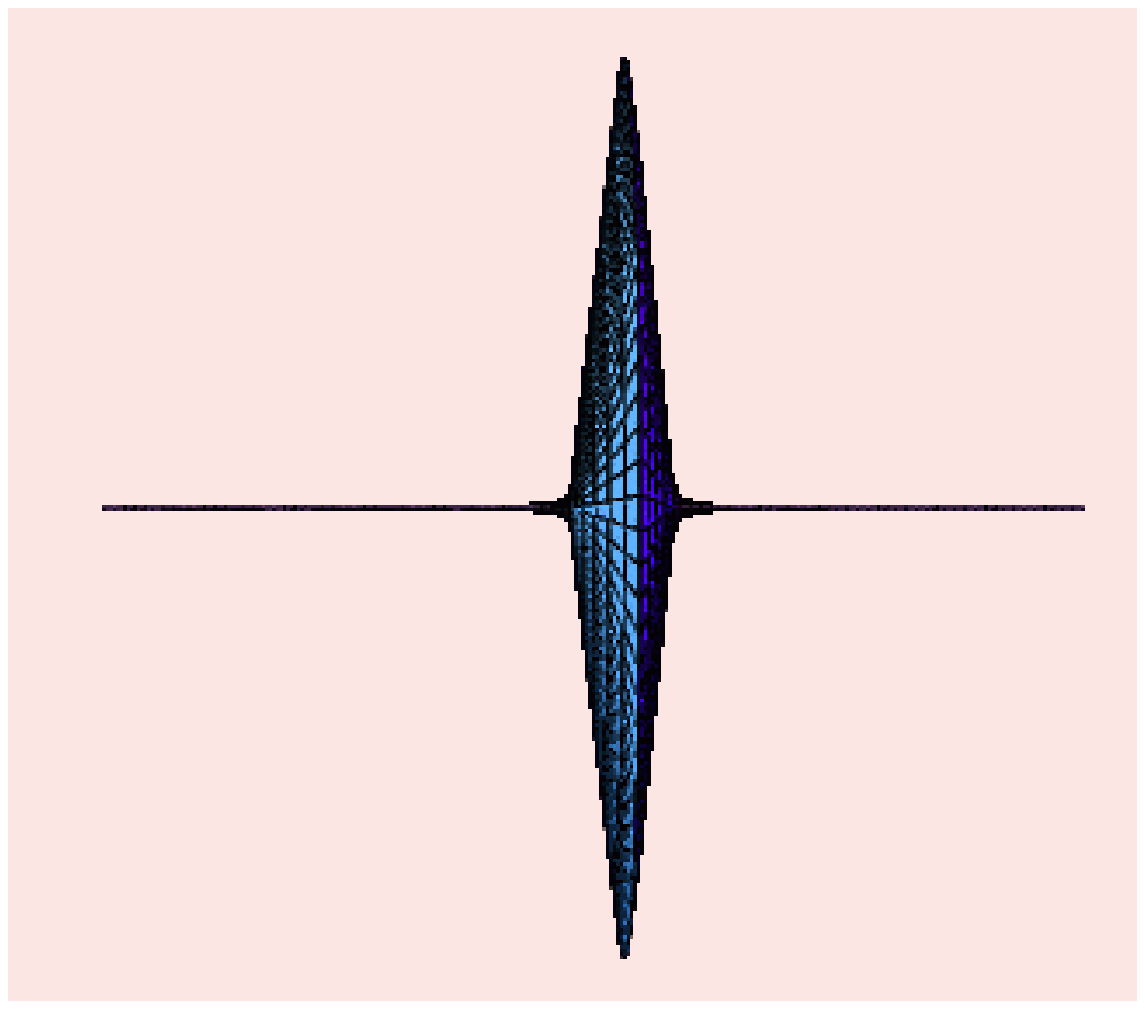}}}}}
\centerline{{\scalebox{0.570}{\rotatebox{90}{\includegraphics{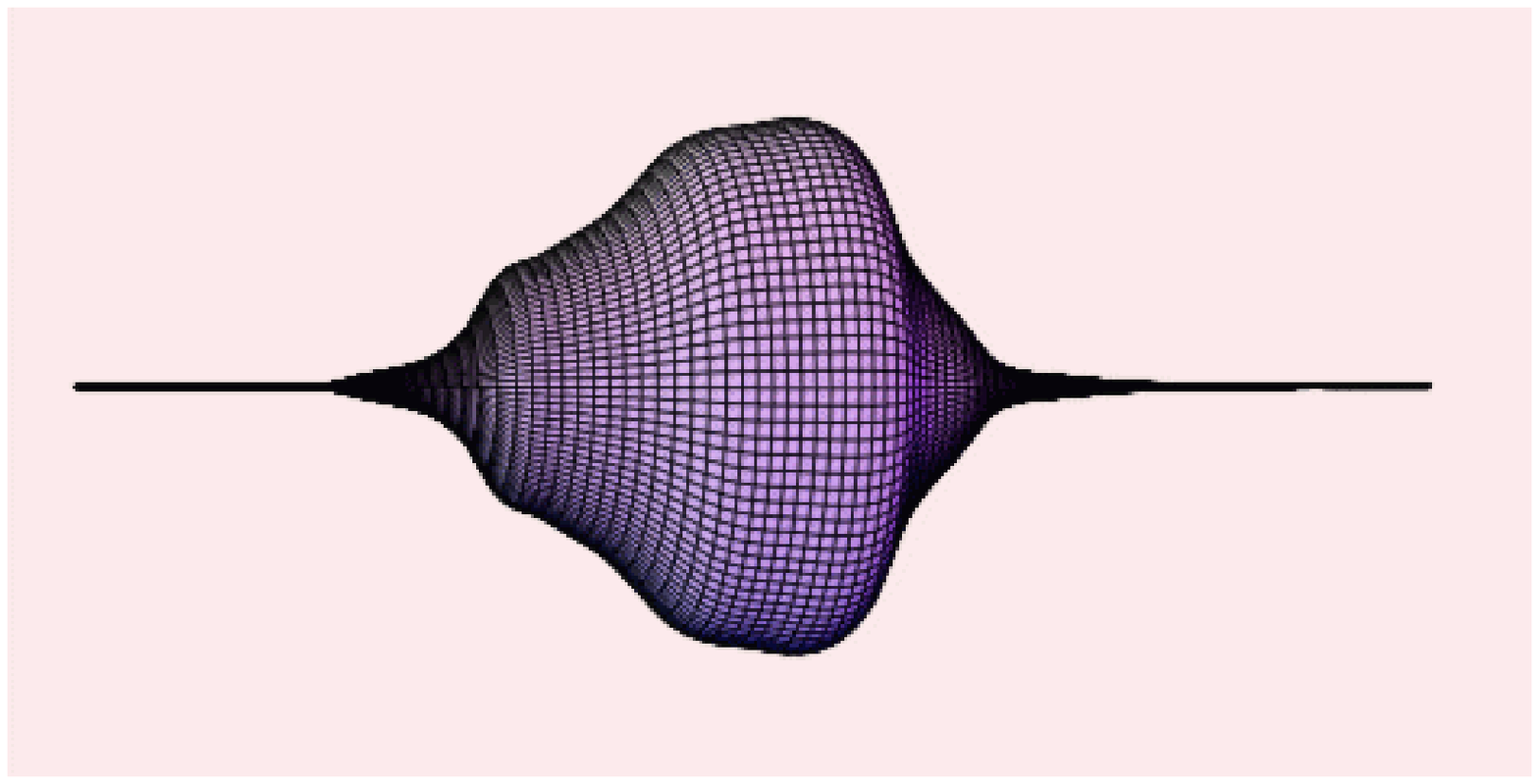}}}}}
\caption{The volume profiles of typical path integral 
configurations in phases A, B and C (no averaging over histories involved).
Phase C (bottom figure) is the one where extended four-dimensional 
geometries emerge. The figures show {\it only} the curves
$V_3(\tau)$ -- made into bodies of
revolution about the horizontal time axis -- and no local fluctuations of 
geometry.}
\label{blobs}
\end{figure}

What is the nature of the extended spacetime found in
phase C of CDT quantum gravity, and what quantitative criteria do we apply
to distinguish between the three phases? 
Examining individual path integral histories will only be of limited use, since
in the limit $a\rightarrow 0$ their geometry will become highly singular, similar 
to that of the nowhere differentiable paths which constitute the carrier space
of the path integral
of the nonrelativistic particle in the continuum limit \cite{reedsimon}. 
What we must do instead is to define and measure
geometric {\it quantum observables}, evaluate their expectation values on the
ensemble of geometries and
draw conclusions about the behaviour of the ``quantum geometry'' generated
by the computer simulations (that is, the ground state of minimal Euclidean
action). 

One such observable is given by the overall shape of the universe, more precisely, 
the three-volume $V_3(\tau)$ as a function of proper time $\tau$. 
Already by comparing
Monte Carlo ``snapshots'' of typical shapes, one observes completely different
qualitative behaviours in the three phases (Fig.\ \ref{blobs}).
Remarkably, inside phase C the microscopic building blocks
superposed in the nonperturbative path integral arrange themselves into an
extended quantum spacetime whose macroscopic shape is that of the well-known
{\it de Sitter universe} \cite{desitter,desitter1}. This amounts to a highly nontrivial
test of the classical limit, about which it is notoriously difficult to make any
definite statements in most models of nonperturbative quantum gravity.
The dynamical mechanism by which this happens is not understood in detail, 
however, it is clear that ``entropy'' (in other words,
the measure of the path integral, or the number of times a given Boltzmann factor
$\exp(-S)$ is realized) plays a crucial role in producing the outcome. 
It means that the nature of this semiclassical limit is 
{\it truly nonperturbative}, in the sense that the tentative continuum limit 
of the path integral 
is found in a region of the bare coupling constant space 
where the entropy of the various geometric configurations contributes at
the same order of magnitude as the exponential of the action. 
As we have pointed out in \cite{semilim}, this is
reminiscent of certain phenomena in condensed matter physics, like
the Kosterlitz-Thouless transition in the XY model.
  
\begin{figure}[t]
\centering
\vspace*{10pt}
\includegraphics[width=9cm]{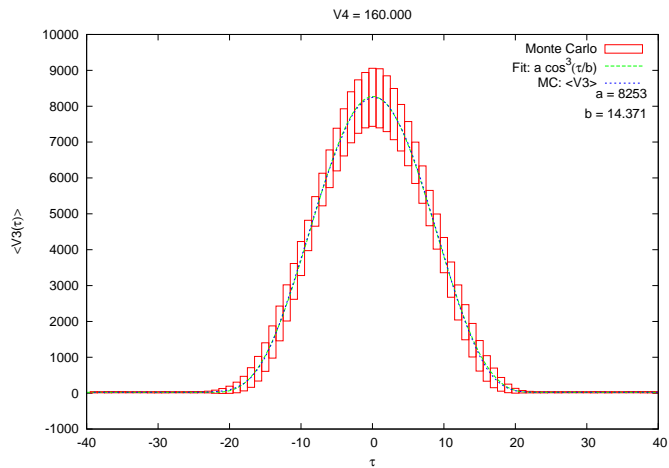}
\vspace*{10pt}
\caption{The average shape $\langle V_3(\tau)\rangle$ of the CDT 
quantum universe in phase C, fitted 
to that of Euclidean de Sitter space (the ``round four-sphere'') with rescaled 
proper time, $\langle V_3(\tau)\rangle=a \cos^3(\tau/b)$. Measurements taken for a universe of 
four-volume $V_4=160.000$ and time extension $T=80$. The fit of the Monte Carlo 
data to the theoretical curve for the given values of $a$ and $b$ is impressive. 
The vertical boxes quantify the typical scale of quantum fluctuations scale around 
$\langle V_3(\tau)\rangle$.}
\label{spherefit}
\end{figure}
The manner in which we have identified (Euclidean) de Sitter space from the
computer data is by looking at the expectation value of the
volume profile $V_3(t)$. 
From the line element of Lorentzian de Sitter space in proper-time coordinates,
\begin{equation}
ds^2=-dt^2+c^2 \cosh^2\left(\frac{t}{c}\right)\; d\Omega_{(3)}^2,
\end{equation}
with $d\Omega_{(3)}^2$ denoting the line element of the unit three-sphere,
one can immediately read off the classical volume profile
\begin{equation}
\label{profile}
V_3(t)= 2 \pi^2 (c \cosh \frac{t}{c})^3, \;\; c=const,
\end{equation}
which for $t>0$ gives rise to the familiar, exponentially expanding universe,
thought to give an accurate description of our own universe at late times,
when matter can be neglected compared with the repulsive
force due to the positive cosmological constant. 
Because the CDT simulations
for technical reasons have to be performed in the Euclidean regime, we
must compare the expectation value of the shape with those of the
analytically continued expression of (\ref{profile}), with respect to the
Euclidean time $\tau:=-i t$. After normalizing the overall four-volume and
adjusting computer proper time by a constant to match continuum proper
time, the average volume profile obtained is depicted in Fig.\ \ref{spherefit}.

A few more things are noteworthy about this result. First, despite the
fact that the discrete CDT construction treats
space and time differently, at least on large scales the full isotropy is restored by the
ground state of the theory for precisely one choice of identifying 
proper time in the continuum. Second, the computer simulations 
necessarily have to be performed for finite, compact spacetimes, which also
means that a specific choice has to be made for the spacetime topology. 
For simplicity, to avoid having to specify boundary conditions, it is usually
chosen to be $S^1\times S^3$, with time compactified\footnote{the period
is chosen much larger than the time extension of the universe and does
no influence the result} and spatial slices which are topological three-spheres.
What is reassuring is the fact that the bias this choice 
could in principle have introduced 
is ``corrected'' by the system, which clearly is driven dynamically to the
topology of a four-sphere (or as close to it as is permitted by the kinematical constraint
imposed on the three-volume, which is not allowed to vanish at any time).
Lastly, we have also analyzed the quantum fluctuations around the
de Sitter background; they match to good accuracy a continuum
saddlepoint calculation in minisuperspace \cite{desitter1}, which is
one more indication that we are indeed on the right track. 

\section{Key findings of CDT -- getting a handle on Planckian physics}\label{sec5}
\begin{figure}[t]
\centerline{\scalebox{0.8}{\rotatebox{0}{\includegraphics{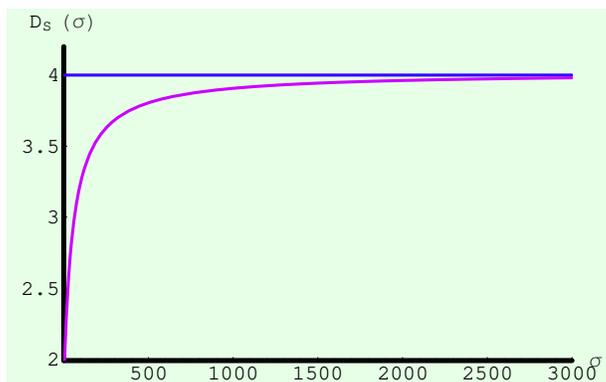}}}}
\caption{The spectral dimension $D_S(\sigma)$ of the CDT-generated 
quantum universe (lower curve, error bars not included), contrasted
with the corresponding curve for a classical spacetime, which for
sufficiently short distances is simply given by the
constant function $D_S(\sigma)=4$.
}
\label{specdimnew}
\end{figure}
Having presented some of the evidence that CDT quantum gravity does
possess the correct classical
limit, let us now turn to the {\it new} physics we are ultimately
after, namely, what happens to gravity and the structure of spacetime at
or near the Planck scale. One way of probing the short-scale 
quantum structure of the universe is by setting up a {\it diffusion process} on the 
ensemble of spacetimes, and studying an associated quantum observable. 
For a classical manifold, it is well known that the speed with which an 
initially localized diffusion process spreads depends on the dimension
of the space. Conversely, given a space $M$ of unknown properties, it can be
assigned a so-called {\it spectral dimension} $D_S$ by studying the
leading-order behaviour of the average return probability ${\cal R}_V(\sigma)$
(of random diffusion
paths on $M$ starting and ending at the same point $x$) as a function of the
diffusion time $\sigma$,
\begin{equation}
{\cal R}_V(\sigma):=\frac{1}{V(M)}\int_M d^dx\ P(x,x;\sigma)\propto 
\frac{1}{\sigma^{D_S/2}},\;\;\;\;\; \sigma \leq V^{2/D_S},
\end{equation}   
where $V(M)$ is the volume of $M$, and $P(x,y;\sigma)$ the solution to
the heat equation on $M$. 
Diffusion processes can be defined on very
general spaces, for example, on fractals, which are partially characterized
by their spectral dimension (usually not an integer, see \cite{fractals}). Relevant
for the application to quantum gravity is that the expectation value
$\langle {\cal R}_V(\sigma)\rangle$ can be measured on the ensemble of
CDT geometries, giving us the spectral dimension of the dynamically
generated quantum universe, with the astonishing result that $D_S(\sigma)$
depends on the linear scale $\sqrt{\sigma}$ probed \cite{spectral}! 
The measurements from CDT quantum gravity, extrapolated
to all values of $\sigma$, lead to the lower curve in Fig.\ \ref{specdimnew}, with
asymptotic values $D_S(0)=1.82\pm 0.25$, signalling highly 
nonclassical behaviour
near the Planck scale, and $D_S(\infty)=4.02\pm 0.1$, which is compatible
with the expected classical behaviour. 
We conclude that the quantum geometry dynamically generated by CDT is
definitely not a classical manifold on short scales. 

What is even more remarkable is the fact that
the same kind of short-scale ``dynamical dimensional reduction'' has been 
found recently in a couple of different quantum field-theoretic 
approaches to quantum gravity, namely, a
nonperturbative renormalization group flow analysis of gravity \cite{spectral-RG}                                  
and the novel Ho\v rava-Lifshitz quantum gravity already mentioned earlier
\cite{spectral-hl}. Whether there is a common underlying reason for this remarkable
coincidence -- which might tell us something deeper about the nature of
quantum gravity -- remains to be understood.
Within the CDT framework, further indications for nonclassicality
at Planckian distances come from measurements of geometric structures in
spatial slices $\tau=const.$ \cite{ajl-rec}, including a measurement of 
their Hausdorff and spectral dimensions, and of shell decompositions of both
space and spacetime \cite{shells}.

\section{Quantum gravity - quo vadis?}

For a long time now, there has been plenty of abstract reasoning on the
nature of nonperturbative 
quantum gravity, that is, what the theory should look like and
what kind of properties it should have {\it if only} we knew what it was.
On the one hand, it is of course
natural to appeal to general principles in the absence of any experimental or observational guidance on how to construct the theory. On the
other hand, our so-called intuition -- mostly coming from studying classical gravity
and quantum fields on a fixed background -- may seriously mislead us
when speculating about the nature of spacetime at the Planck scale.
What lattice quantum gravity (in the form of dynamical triangulations or
causal dynamical triangulations) provides us with is an ``experimental lab'',
a calculational framework to study systems of fluctuating geometry
{\it quantitatively} in a nonperturbative regime. In dimension two, where
comparisons with analytical models are available, this leads
to sensible results. In dimension four, it is currently the only way to extract
nonperturbative information about these systems. In particular, it has uncovered
several completely unexpected, but presumably generic features, for example, 
the fact that the signature of the geometry can make a crucial difference,
the fact that a superposition of $d$-dimensional geometries is not
necessarily $d$-dimensional, indeed, that such superpositions are usually
so degenerate that they possess no classical limit at all, and the fact that 
the conformal divergence of the Euclidean path integral can be cured by ``entropic
contributions''.  

CDT's toolbox has enabled us to uncover these nonperturbative properties
and at the same time make quantitative
statements about covariant properties of quantum geometry, including
its dimension and volume profile. In principle the framework is also 
able to test nonperturbative predictions from other fundamental theories
containing gravity, if and when they will be made,
subject only to the usual numerical limitations of the lattice.
Clearly, much remains to be done, but the results already obtained underline
the power and utility of lattice methods, also in situations where spacetime
itself is dynamical.

\end{document}